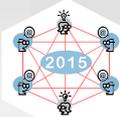

# Looking into Hardware-in-the-Loop Coupling of OMNeT++ and RoSeNet

Sebastian Boehm and Michael Kirsche
Computer Networks and Communication Systems Group
Brandenburg University of Technology Cottbus-Senftenberg, Germany
eMail: {sebastian.boehm, michael.kirsche}@b-tu.de

*Abstract*—Network emulation using real sensor node hardware is used to increase the accuracy of pure network simulations. Coupling OMNeT++ with network emulation platforms and tools introduces new application possibilities for both sides. This work-in-progress report covers our experiences of using OMNeT++ as a test driver for RoSeNet, a network emulation and test platform for low-power wireless technologies like IEEE 802.15.4. OMNeT++ and RoSeNet were interconnected to enable a co-simulation of real sensor networks with a MAC layer simulation model. Experiences and insights on this Hardware-in-the-Loop (HIL) simulation together with ideas to extend OMNeT++ and to provide a generic interconnection API complete the report.

## I. INTRODUCTION

The evaluation of wireless networks requires accurate and extensive testing using tools and concepts of network testbeds such as packet, routing and latency analyses. Due to the underlying conditions in the wireless spectrum, a node's energy consumption and performance aspects also need to be considered. The performance and features of distributed sensor applications strongly depend on the underlying hardware resources and the channel conditions. This performance impact can usually only be tested and debugged on the actual sensor node hardware. With an increasing number of nodes in a network, installing and configuring application-specific testbeds is a time consuming and often expensive task.

Different crucial network evaluation techniques for Wireless Sensor Networks (WSNs) are presented below. We emphasize the necessity of network emulation in general and motivate our contribution of coupling OMNeT++-based network simulation with emulation tools like RoSeNet in context of WSNs.

### A. Simulation of Wireless Sensor Networks

Within network simulations, it is relatively easy to define network topologies, layered architectures, and algorithmic applications. In the simulation flow, where system states change at discrete points over time, parameters can be customized individually to determine performance factors. In the context of WSNs however, simulations are limited [1] and lack realism. Simulators in general may not provide accurate modeling of real execution times at node level. Nodes are commonly simple and abstract entities, where actual hardware resources do not have any representation. Suitable radio channel and physical environment conditions are typically abstracted or omitted completely; both have to be considered when developing applications for resource constrained devices working in wireless environments. Wireless environments and channel conditions are depicted and abstracted by mathematical functions from varying complexity. There is a big tradeoff between realistic simulation models and their performance and scalability. A simple example is the comparison between the performance and realism of wireless channel models like the disk or free space model, the log normal shadowing model, and the multi-path ray tracing model. Their performance is decreasing while their correctness is increasing (although at different rates).

### B. Wireless Sensor Network Testbeds

For the case of distributed sensor networks, it is usually complex and costly to manage physical testbeds with all required analysis features. Nodes within testbeds are not isolated from their surrounding radio environment, which is difficult to predict and measure. Common physical WSN testbeds lack controllability and easy configuration of the physical environment and are often limited or fixed in scale. In general such testbeds cannot enable easy adjustments of application or protocol parameters as it is common in network simulators. Applications run as static implementations on sensor nodes, which is why live debugging and detailed monitoring of system states and traces during the application execution are often not provided inside testbeds. Firmware extensions and extended runtime environments with monitoring capabilities are options to provide such information in testbeds. However, their maintenance is time-consuming and interventions in the running implementation can introduce unforeseen behavior and errors. Common WSN experimentation testbeds are presented and compared in [2].

### C. Wireless Sensor Network Emulators

Controllable operational and environmental system conditions for physical signal transmissions are feasible in hardware-based network emulation. The network emulation concept forms an entire evaluation system through a combination of various real, virtual and abstract components. In the context of developing a sensor network application, the actual target hardware is still essential because both the PHY and (partially) the MAC layer are implemented in hardware and hardware constraints play an important role in application and protocol design. An interesting evaluation technique is provided by wireless channel emulators. Practical deployments vary from laboratory test setups with coaxial-based radio links [3] to complex analog or digital radio channel emulators [4, 5]. In such test setups, real hardware nodes are shielded and connected over their radio interfaces to a hardware channel simulator, where effects of signal propagation are emulated. In the context of current wireless spectrum dynamics, it could be very important to test node networks and wireless applications under detailed wireless influences. Nevertheless, the traffic in such emulation and test setups will be created by sensor nodes running custom applications and static protocol implementations, comparable to testbeds.

To transcend the limitations of both, simulation and wireless emulation systems, our suggestion is to combine both techniques in a kind of co-simulation, which provides a runtime environment to conduct wireless network experiments under realistic channel conditions by using real WSN hardware including their transceivers. Early survey reports like [6] showed that there are open research questions regarding the integration of tools to support both simulation and emulation. We want to encourage interest in this topic and stimulate discussions with other OMNeT++ community members, to refine the coupling of real hardware with OMNeT++ in the context of WSN simulation.

This work-in-progress report covers our experiences using OMNeT++ in context of hardware-based network emulation with RoSeNet. It is structured as follows: In section II, we introduce RoSeNet, its emulation concept, and our current setup that uses OMNeT++ as a test driver. We also introduce an existing co-simulation approach and an accurate simulation model for IEEE 802.15.4, which is a common PHY / MAC standard in the domain of WSNs. While using OMNeT++ in our test case as the preliminary stage of coupling, section III presents our strategy for a seamless combination of OMNeT++ and RoSeNet on a protocol level. In section IV, important aspects and the main benefits of our contribution are presented. Section V closes this report with a short discussion about the benefits of a generic co-simulation interface for OMNeT++.

## II. BACKGROUND

Our motivation to combine network simulation with emulation tools derives from our main field of research (i.e., network emulation) and our previous work with the network emulation platform RoSeNet. This section covers (a) our setup of using OMNeT++ as a test driver, similar to the co-simulation strategy used by Veins [7], and (b) an accurate IEEE 802.15.4 simulation model [8], necessary for simulation-based network emulation from our point of view.

### A. RoSeNet

RoSeNet [9] is a network emulation and test platform for low-power wireless technologies and WSNs, developed by dresden elektronik[1], which focuses on hardware-based channel emulation. The platform consists of individual panels with replaceable sensor node hardware, including custom application and protocol implementations. The overall architecture incorporates up to 1000 sensor nodes, which enables emulation of large-scale networks. All sensor nodes on the platform are interconnected through a controllable coaxial cable radio environment. RoSeNet can hence be described as a wireless channel emulator where signal propagation can be adjusted by digital step attenuation. Interference signals can be injected with the help of signal supply points. The status of RoSeNet can be considered as "in development".

During our work with the RoSeNet network emulator, we developed an interface that enables us to "feed" real sensor node hardware on the emulation panels with generated MAC layer protocol data. Developers can thus achieve a decent control over the communication flow in the network and construct various test scenarios. To generate traffic for the RoSeNet emulation platform, we inject frames using OMNeT++ as a test driver. We created an example in OMNeT++ / INET to feed our emulation setup with generated MAC protocol data units. For a test case, we simulated a scenario with TCP/IP traffic over Ethernet. We hence transmitted non-compliant protocol data to the destination hardware over the Hardware-in-the-Loop (HIL) interface in this "first step" scenario.

To overcome difficulties in creating external sensor node interfaces in OMNeT++, we used and extended INET's `PcapRecorder`. PCAP is a widely used standard for the exchange of communication data over program and system boundaries. Because PCAP frames already include real binary packet formats for most of the popular and currently used protocols, we have a common and widely used standard for exchanging our communication data. We implemented changes to utilize `PcapRecorder` together with a network dumper. Within the initialization, the `PcapRecorder` establishes a TCP connection with the `PcapDumper` module to the emulation server of the RoSeNet system. The parameters of the socket connection are set in the simulation configuration file (compare Listing 1).

Listing 1. Socket connection configuration - excerpt from `omnetpp.ini`
```
**.server.numPcapRec = 1
**.server.pcapRecorder[0].networkEnabled = true
**.server.pcapRecorder[0].serverIP = "localhost"
**.server.pcapRecorder[0].serverPort = 4242
**.server.pcapRecorder[0].pcapFile = "results/server.pcap"
```

Simulated packets can now be transmitted through the socket connection during simulation runtime. Inside RoSeNet's software system, these packets are interpreted, encapsulated, and forwarded to the specified nodes over their serial interface via the Serial Line Internet Protocol (SLIP). Incoming protocol data packets at node level are transmitted over the network stack. On the return path, radio packets will be forwarded to RoSeNet's software system, encapsulated in the PCAP format and transmitted to any registered receiver, even back to the simulation system (OMNeT++ in our case). The flow chart in Figure 1 only depicts a simplified forwarding of packet data from OMNeT++ via the sub-modules of the RoSeNet emulation system to the radio hardware.

---
[1]RoSeNet research project: https://www.dresden-elektronik.de/funktechnik/wireless/research-projects/rosenet/?L=1

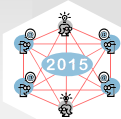

## C. IEEE 802.15.4 Simulation Model

We focused on a new IEEE 802.15.4 simulation model for OMNeT++ / INET that was previously introduced at the OMNeT++ Community Summit in 2014 [8]. The simulation model was built to emulate the complex behavior of the IEEE 802.15.4 standard in the 2006 revision. Protocol operations and service primitives were designed to be compliant with the standard. A detailed introduction of the simulation model is available in [8]. Figure 2 shows an IEEE 802.15.4 example host with a network stack consisting of different submodules.

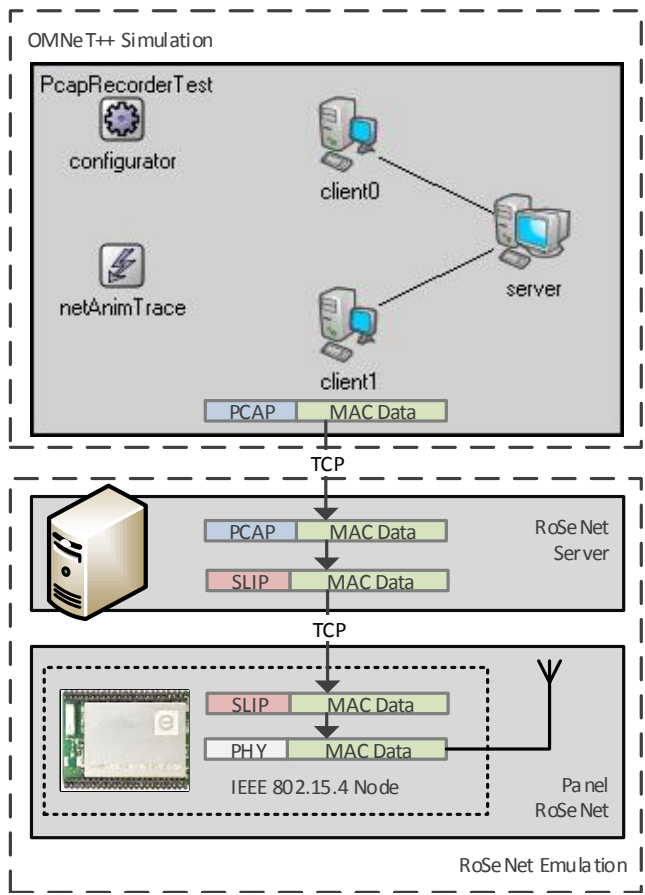

Fig. 1. Simplified message exchange between OMNeT++ and RoSeNet

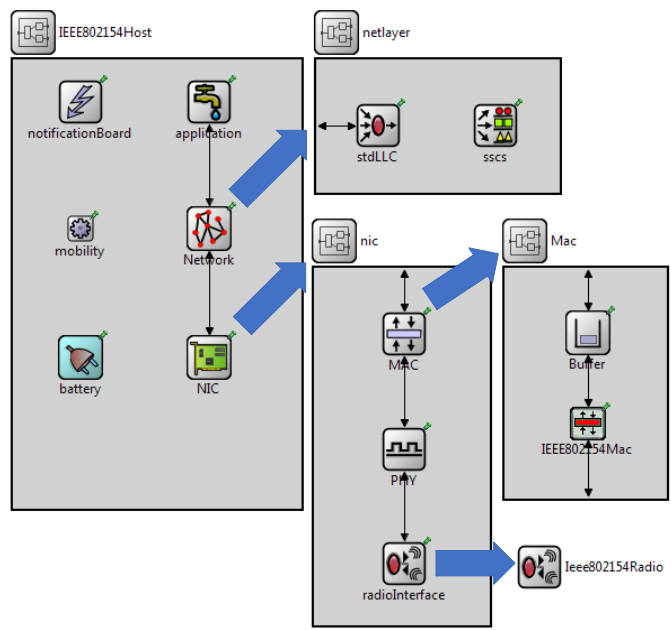

Fig. 2. An IEEE 802.15.4 host with its submodules (taken from [8])

### B. Co-Simulation in OMNeT++

We motivated the need for a simulation of both physical and communication network dynamics of WSNs in Section I. A comparable concept for a co-simulation of OMNeT++ with another tool (software tool in this case) can be found in the area of mobility simulation, which involves highly dynamic and complex systems. Actor mobility is a major concern in the research field of Vehicular Ad Hoc Networks (VANETs). The Veins [7] framework uses a co-simulation concept for a bidirectional coupling of OMNeT++ and the SUMO (Simulation of Urban MObility) traffic simulator. Coupling and communication are controlled and synchronized over the Traffic Control Interface (TraCI) [10]. TraCI uses a TCP socket connection to enable a bidirectional exchange of command and response messages between both Veins (OMNeT++) and SUMO. The co-simulation concept of Veins and SUMO is based on the processing of internal and external events, whereupon OMNeT++ retains control over the simulation flow.

Retaining control of RoSeNet's software and runtime system for test scenarios through the use of OMNeT++ is also our motivation. We hence need an accurate representation of the IEEE 802.15.4 standard in OMNeT++ / INET, to exchange the necessary protocol primitives with the real hardware transceiver implementations.

## III. SIMULATION-BASED NETWORK EMULATION

In this section, we introduce and motivate a strategy that combines model-based network simulations and real sensor hardware to increase the accuracy of WSN evaluation. Some parts in this combination strategy, particularly the higher layers of the protocol stack, are simulated, while other parts are represented by real hardware components interconnected through HIL interfaces. For our own work, we want to extend OMNeT++'s capabilities to enable physical channel emulation and hardware-related profiling of protocol performance characteristics in sensor networks through the exchange of protocol and control data units among OMNeT++ and RoSeNet. This concept of bridging between simulated and physical real nodes could speed up and expand the possibilities of wireless embedded system design. Other application areas would also benefit from such a coupling of OMNeT++ with external tools over a desired common interface.

We argue that the interconnection of network simulation and network emulation platforms and tools, where real physical wireless network and sensor interfaces are used, could be a valuable extension for an OMNeT++-based network analysis and performance evaluation toolset. The exchange of real and simulated protocol data can enhance accuracy in a similar way that a preparation of simulations with physical sensor readings and protocol data could improve the exactness of simulations when compared to testbeds. Network emulators, on the other hand, could also be fed with simulated protocol data to verify and validate used simulation models. Such interfaces are still not included in the area of WSN simulation.

Our suggestion is a hybrid network emulation system in a HIL or emulation-assisted simulation environment, where the emulation setup is controlled by the network simulation. In practical terms, OMNeT++ exercises control over the sensor hardware or the hardware emulation system. To enable this, extensions of the simulation framework are planned to support parallel execution of protocol simulation and physical packet transmission. The proposed approach allows to include the physical details (e.g., radio transceiver or sensor interfaces) of a sensor node in protocol simulations, which are very costly to model but often feasible in testbeds or network emulation platforms like RoSeNet. To avoid being reliant on a specific node platform, we focused on a common interface that includes mechanisms for sending and receiving MAC packets as well as service primitives conform to the IEEE 802.15.4 standard over the boundaries of the simulation domain. Our contribution is also to provide OMNeT++ with control interfaces for acquiring data from different sensor node hardware or emulation systems.

*A. HIL Interface Architecture*

The resulting co-simulation interface is responsible for controlling interactions and the communication flow between simulated and real or emulated nodes. Developers could replace some simulated nodes within their scenarios with physical ones to verify hardware-depending features. Furthermore, we do not assume that both evaluation domains run on the same host system. The synchronization and control scheme between simulator and real sensor hardware could be realized similar to the Veins approach, focused on controlling the hardware runtime system by the simulation. Therefore, a number of components are necessary to generate emulation parameters, exchange messages in real packet format, and manage a synchronized real-time communication and control flow. To facilitate a consistent and universal communication interface solution, we focus on exchanging protocol data at the MAC layer using PCAP, like we first did while using OMNeT++ / INET as a simple test driver.

*1) External Interfaces:* In context of IEEE 802.15.4 and the used simulation model [8], the basic component in OMNeT++ / INET is represented as a protocol-specific external interface (e.g. `IEEE802154ExtInterface`) that handles communication data as events from both abstract simulation and emulation domain. A structural architecture of this external interface could be comparable to Irene Ruengeler's approach, where INET is connected with real networks [11, 12]. Figure 3 shows an example external interface for IEEE 802.15.4. Protocol-conform serializers and de-serializers will be used to convert simulated protocol data to the real binary packet format and back. Receiving and forwarding of packets needs to be implemented by a custom PCAP scheduler with access to the HIL interface. To enable the control of a node's real physical interface from within OMNeT++, we have to identify possible exchange points for the transfer of protocol messages between simulated and real layers. Those exchange points could be implemented in the node's firmware or its operating system.

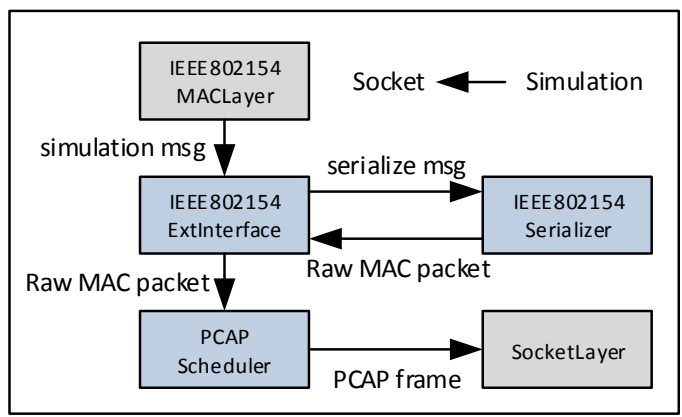

Fig. 3. Example IEEE 802.15.4 external interface in OMNeT++ / INET

*2) Emulation Control Interface:* The communication and control flow of the HIL approach is ensured over an Emulation Control Interface (EmuCI), similar to the TraCI approach. A connection to the emulator will be realized over a TCP socket in conjunction with custom schedulers that take responsibility for communication and control flows. The interface needs to provide all necessary mechanisms to control the emulator and all emulation-characteristic parameters. The inter-domain transmission of simulated and real packets can be realized with a communication-related PCAP scheduler. Parameters and settings should be included in the simulation configuration process to simplify things.

IV. SCENARIOS AND BENEFITS

In the context of network evaluation in realistic environments, the objective is to couple the event-based network simulation with network emulation tools or real sensor node hardware through an interface that fulfills the requirements of a realistic evaluation using hardware-based emulation. This way, it is possible to control the behavior of emulated WSN environments to understand the influences of wireless channels and hardware characteristics from an abstract network simulation's point of view. During simulation runtime, adaptations to external real or emulated components could be enforced.

Important aspects for a validation of radio transmissions and wireless mediums are not integrated in available simulators for wireless sensor networks. A few desired benefits of the simulation-controlled emulation concept are:

- the ability to feed a WSN simulation with real application data and sensor readings,
- feed the sensor node hardware in context of implemented test applications with simulated traffic and network data,

- enable a power profiling of communication overhead at the PHY layer,
- wireless channel emulation support for simulation, e.g. injecting interferences and non-compliant protocol data or jamming the radio communication.

## V. Discussion

These contributions shall motivate discussions about the benefits of a generic interface for co-simulation architectures. From our point of view, a generic co-simulation interface instead of an application- or framework-specific one is an interesting extension for OMNeT++. With Internet of Things scenarios, VANETs, intelligent sensor networks for Smart Grid or medical applications, and Cyber Physical Systems (CPSs) interacting with cloud services, there is a growing demand to simulate hardware or environment-related details and network protocol behavior simultaneously in a comprehensive manner. The exchange of system events on various levels is an objective. The event-driven concept behind network simulation with OMNeT++ would be preserved and extended with external events. Communication-specific details and protocol message flow should be allowed at different protocol layers.

## VI. Final Remarks

We want to state that the presented approach is work-in-progress, thus still in a high-level state without fully laid out technical details. The conceptual design behind our ongoing research and the proposed ideas should be not limited to any specific emulation system, hardware resource, or network simulation environment. We will however focus on OMNeT++ and RoSeNet for the actual implementation and testing.